\documentclass[
twocolumn,
prd,amsfonts,eqsecnum,nofootinbib,amsmath,showpacs,superscriptaddress
]{revtex4}

\usepackage{bm,graphics,graphicx,epsfig,color,ulem}

\begin{document}
\title{Synergy of short gamma ray burst and gravitational wave observations: Constraining the inclination angle of the binary and possible implications for off-axis gamma ray bursts}
\author{K. G. Arun}\email{kgarun@cmi.ac.in}
\affiliation{Chennai Mathematical Institute, Siruseri, Tamilnadu, 603103, India}
\author{Hideyuki Tagoshi}\email{tagoshi@vega.ess.sci.osaka-u.ac.jp}
\affiliation{Department of Earth and Space Science, Graduate School of Science, Osaka University, Osaka, 560-0043, Japan}
\author{Archana Pai}\email{archana@iisertvm.ac.in}
\affiliation{Indian Institute of Science Education and Research Thiruvananthapuram, Computer Science Building, College of Engineering Campus, Trivandrum, Kerala, 695016, India}
\author{Chandra Kant Mishra}\email{chandra@icts.res.in}
\altaffiliation[Presently at]{ the International Centre for Theoretical Sciences, Tata Institute of Fundamental Research, Bangalore 560012, India}
\affiliation{Indian Institute of Science Education and Research Thiruvananthapuram, Computer Science Building, College of Engineering Campus, Trivandrum, Kerala, 695016, India} 
\date{\today}
\begin{abstract}
Compact binary mergers are the strongest candidates for the progenitors of Short Gamma Ray Bursts (SGRBs). If a gravitational wave (GW) signal from the compact binary merger is observed in association with a SGRB, such a synergy can help us understand many interesting aspects of these bursts.  
We examine the accuracies with which a world wide network of gravitational wave interferometers would measure the inclination angle (the angle between the angular momentum axis of the binary and the observer's line of sight) of the binary. We compare the projected accuracies of GW detectors to measure the inclination angle of double neutron star (DNS) and neutron star-black hole (NS-BH) binaries for different astrophysical scenarios.
We find that a 5 detector network can measure the inclination angle to an accuracy of $\sim 5.1 (2.2)$ degrees for a DNS(NS-BH) system at 200 Mpc if the direction of the source as well as the redshift is known electromagnetically.
We argue as to how an accurate estimation of the inclination angle of the binary can prove to be crucial in understanding off-axis GRBs, the dynamics and the energetics of their jets, and  help the searches for (possible) orphan afterglows of the SGRBs.
\end{abstract}
\maketitle
\section{Introduction}
\subsection{Gamma Ray Bursts}
Gamma Ray Bursts (GRBs) are classified as `long' and `short' based on the duration ($T_{90}$) in which 90\% of the total observed energy is emitted~\cite{ShGRBclass93} and  the spectral type. Long duration GRBs correspond to $T_{90}>2 s$ and are spectrally soft whereas the short GRBs (SGRBs henceforth) correspond to $T_{90}<2s$  and are spectrally hard. Long duration GRBs occur at relatively high redshifts and in active star forming regions in the Galaxy and many of them have been associated with core-collapse supernovae~\cite{Hjorth2003dh}. On the other hand, the progenitors of SGRBs are not fully understood. They have relatively smaller redshifts 
 and are seen with significant (tens of kpcs) off-sets with respect to the respective Galactic centers~\cite{FongBerger2013,ChurchEtal2011}. The inferred progenitor ages are typically a few Gyr pointing to older stellar populations~\cite{LeiblerBerger2010}. These features strongly support the hypothesis that they are due to the mergers of double neutron stars (DNS) or neutron star-black hole (NS-BH) binaries \cite{EichlerShGRB89,Narayan92}. Recent kilonova observation associated with the GRB130603B~\cite{KilonovaBerger2013,TanvirKilonova2013} reinforces this conjecture (see Refs.~\cite{ShGRBrevLee07,Nakar07Review,ShGRBrevBartos2013,BergerShGRBrev2013} for reviews on SGRBs and their association with gravitational waves (GWs).).

The prompt emission and afterglows of the SGRBs share a lot of features of the long GRBs. This is because for both the bursts, the  central engine is a black hole that accretes from the surroundings, powering a jet  which produces prompt emission due to internal shocks and the afterglow by its interaction with the circumburst medium.\footnote{In the case of SGRBs, there may also be an intermediate product in the form of a hypermassive, highly magnetized neutron star which eventually collapses to a BH~\cite{RosswogEtal03,AloyEtal05,ShibataEtal05,Giacomazzo2010,Hotokezaka2011,Hotokezaka2013}.}  Hence the standard fireball model~\cite{Rhoads99,SariEtal99}, which has been very successful in interpreting the long GRBs  is used to interpret the SGRB data as well.  
 But the central black hole for the these two bursts is formed via different channels. For long GRBs, the BH is formed by core collapse of a massive star, whereas for the SGRBs they may be formed by
mergers of compact binaries.
\subsection{GRB-GW association}
If indeed the SGRBs are produced due to the mergers of compact binaries, there will also be an associated emission of GWs. If the burst happens sufficiently close-by, the corresponding GW signals may be detectable by the upcoming advanced GW interferometers such as advanced LIGO (aLIGO)\cite{aligo}, advanced Virgo~\cite{avirgo} and KAGRA~\cite{kagra}. The expected detection rate at design sensitivity for aLIGO is approximately 40(10) per year for DNS(NS-BH) sources~\cite{LSCrates}. The {\it root mean square} distance reach~\footnote{This should be distinguished from the horizon distance which is the distance to an optimally oriented binary. $D_{\rm horizon}=\frac{5}{2}D_{\rm rms}$  (see Eq.~(4.12) of \cite{DIS00}).} for DNS binaries with a signal to noise ratio (SNR) of 8 for aLIGO detector is $\simeq200$ Mpc ($z\simeq 0.05$) whereas for the NS-BH system (total mass $11 M_{\odot}$), it is roughly $1$ Gpc ($z\simeq 0.2$)(see Fig. 1 of ~\cite{AISS05} and rescale it to a SNR of 8).  This distance reach will increase with the number of detectors ($n$) in the network, roughly as $\sqrt{n}$. Hence, for a 5 detector network the distance reach can be, roughly, twice the one given above. Hence the advanced GW detector era 
carries the exciting prospect of detecting GW signals from compact binary mergers and detection of associated SGRBs (and their afterglows) by  electromagnetic (EM) telescopes.  Typical numbers of joint SGRB-GW events were estimated to be $\simeq 0.2-1(1-3) {\rm yr^{-1}}$ for a DNS(NS-BH) progenitor~\cite{DietzEtal2013}. 

This opens up lot of interesting studies that are possible where the synergy of GW and EM observations may lead to significant breakthroughs in the field of astronomy, specifically related to SGRBs.   Such an association would confirm, beyond question, the compact binaries as progenitors of the SGRBs. GW observations would measure the  (redshifted) masses of the binary to a few percent accuracy~\cite{AISS05} which will further enable us to distinguish between the DNS and NS-BH scenarios. One of the most exciting possibilities of such joint GW-EM observation is the estimation of Hubble constant without relying on the usual astronomical distance ladders, as suggested by Schutz~\cite{Schutz86} and explored in greater detail in Refs.~\cite{DaHHJ06,Nissanke09a}(see, also Refs.~\cite{,SathyaSchutzLivRev09,ShGRBrevBartos2013,BergerShGRBrev2013} for an overview of other ideas proposed in the literature). In this paper, we investigate a different aspect related to the possible GW measurement of the inclination angle (the angle between the line of sight of the observer and the angular momentum axis of the binary), 
(See also an another discussion by Chen and Holz ~\cite{ChenHolz2012} on the relation between jet opening angles 
of the SGRBs and the detection rate of NS-NS and NS-BH binary mergers in the advanced LIGO era). 

\begin{figure}\includegraphics[scale=0.4]{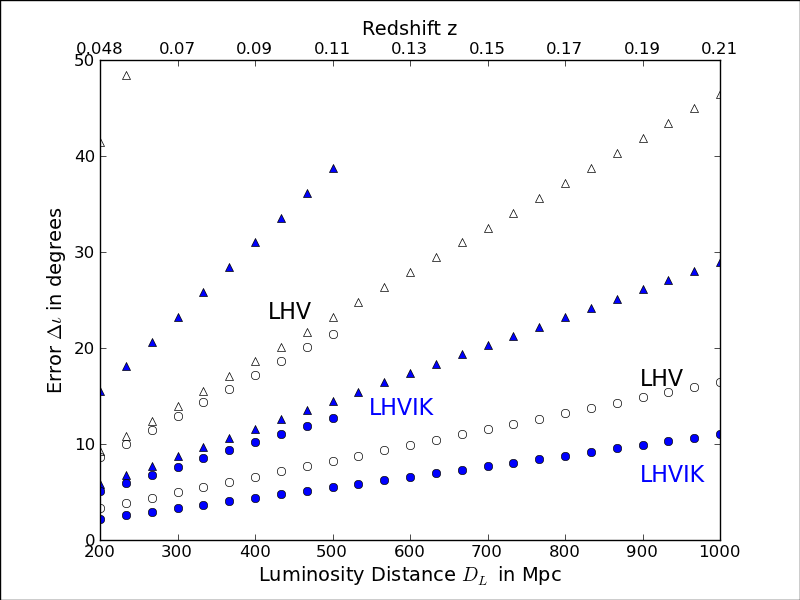}
\caption{Estimated median errors in the measurement of the inclination angle of the binary (also the viewing angle of the SGRB jet) as a function of luminosity distance (bottom axis) and redshift (top axis) for different detector configurations, different astrophysical scenarios, and for DNS and NS-BH binaries. The triangles correspond to the astrophysical scenario where there is no EM information, and the circles correspond to 3D localized SGRBs (see text for details). The filled and open data points distinguish the 3 detector GW detector configuration LHV (two advanced LIGO detectors and one advanced Virgo) and a 5 detector configuration LHVIK (adding Japanese detector KAGRA and the proposed LIGO-India to LHV). Errors for the DNSs are characterized by the lines which are cut at 500 Mpc and those of NS-BH are characterized by those truncated at 1 Gpc. We have not quoted errors greater than 50 degrees in this plot as it may not be of any astrophysical relevance. }\label{fig:summary}

\end{figure}

We discuss the accuracies with which advanced GW detectors would be able to measure the inclination angle of the binary for various astrophysical scenarios.   The inclination angle is referred to as the {\it viewing angle} of the jet (angle between the jet axis and the line of sight of the observer) in the GRB literature,  assuming the jets are launched along the orbital angular momentum axis, which is the same as the total angular momentum axis for nonspinning binaries or for binaries whose spins are aligned or anti-aligned with respect to the angular momentum axis. The viewing angle can play an important role in GRB modelling if the jets are pointed away from the observer (off-axis jets).  We discuss how the SGRB-GW synergy can shed light on the  geometry and energetics of the  SGRBs
 A quick summary of the results are presented in Fig.~\ref{fig:summary} where the expected accuracy with which the inclination angle of the binary may be estimated  for DNS and NS-BH progenitors using GW measurements for various GW detector combinations.

The paper is organized as follows. In Sec.~\ref{sec:model} we discuss the data analysis technique employed in GW astronomy and the model of gravitational waveform we employ for our study. Section~\ref{sec:iota} discusses the accuracies with which GW observation will be able to extract the inclination angle of the binary. The implications of these measurements for SGRBs are discussed in Sec.~\ref{sec:GRB} and caveats of our model and future plans are discussed in Sec.~\ref{sec:caveats}.
\section{Gravitational waveform model}\label{sec:model} 
Matched filtering, a well-known phase matching technique in data analysis, will be used to detect GW signals from compact binaries characterized by various parameters (masses, spins, luminosity distance, source location, inclination etc.) in noisy data. Matched filtering makes the best use of the prior predictability of the gravitational waveforms using General Relativity, the most successful theory of gravity to date. The gravitational waveforms produced during the entire evolution of the binary,  are obtained by using various analytical or numerical approximation techniques such as post-Newtonian (PN) theory~\cite{Bliving} for the adiabatic inspiral phase, numerical relativity~\cite{Pretorius07Review} for the strong field merger phase, and black hole perturbation theory for the ringdown phase~\cite{TSLivRev03}.

 For low mass systems, the adiabatic inspiral phase contributes the most to the waveform that can be detected by the GW interferometers using matched filtering. PN theory allows computation of very accurate waveforms for the adiabatic inspiral phase which ends at the last stable orbit of the binary, when the strong field effects take over. In this work, we use the (point particle results of) PN model of  the binary evolution to model GW signals from such systems. In doing so, we are neglecting the finite size effects which show up in the PN theory very close to the merger, such as those given in \cite{1PNTidal2011} which are higher order PN effects. 
Since frequency domain is a natural basis for performing matched filtering, it is convenient to write down the gravitational waveform from a compact binary system in the frequency domain. The most accurate waveform is one in which the corrections to the phase as well as amplitude of the waveform are accounted to the maximum possible accuracy in the PN theory. This can be obtained starting from the time domain waveforms computed in Refs.~\cite{BIWW96,ABIQ04,BFIS08}, which can be schematically written as
\begin{equation}
h(t)=\frac{ M \eta}{D_L} \sum_{k=1}^{7}\sum_{n=0}^5 A_{(k,n/2)}
v^{n+2}(t)\;\cos\left[k\phi(t)+\phi_{(k,n/2)}\right],\label{eq:FWF}
\end{equation}
where $M$ is the total mass of the binary, and $\eta=\mu/M$ is the symmetric mass ratio (defined as the ratio of reduced mass to the total mass). In addition to the masses, the amplitude $A_{(k,n/2)}$ is a function of the inclination angle of the binary ($\iota$), polar angles specifying the
source location with respect to the detector ($\theta, \phi$) and the polarisation angle ($\psi$). The waveform is an expansion in powers of a dimensionless velocity variable $v=(\pi M f/k)^{1/3}$ and $k$ keeps track of different harmonics of the orbital phase which appears in the waveform, and $\phi_{(k,n/2)}$ denote the phase off-sets of the different harmonics with respect to the leading harmonic at twice the orbital phase of the binary. In the above equation $n$ denotes the PN order ($n=2$ means 1PN).  In this work we have used the waveform which has 2.5PN accurate amplitude~\cite{ABIQ04,ChrisAnand06} and 3.5PN accurate phasing~\cite{BDEI04,DIS02}.
The corresponding frequency domain waveform can be straightforwardly obtained using the stationary phase approximation following the prescription given in Refs.~\cite{ChrisAnand06,ABFO08}. 
\section{Measurement of inclination angle of the binary from GW observations}\label{sec:iota}
 As discussed above, the amplitude of the signal depends on the inclination angle of the binary and this dependence enters the waveform through amplitude functions, $A_{(k,n/2)}$, appearing 
in Eq.~\eqref{eq:FWF}. 
The inclination angle is the angle between the observer's line of sight and the angular momentum vector of the binary. Hence, in principle, using a network of
gravitational interferometers, and using the matched filtering technique, GW observations will be able to extract the inclination angle from the  detected gravitational waveform. 
Next, we discuss the accuracy with which GW observations will be able to determine the inclination angle of the binary.

We use Fisher information matrix formalism~\cite{Rao45,Cramer46,CF94} to obtain the accuracy with which 
the errors on the cosine of the inclination angle ($\cos \iota$) can be estimated. The Fisher information matrix approximates the likelihood function associated with a signal with a multi-variate Gaussian of dimension equal to the number of parameters that need to be extracted, which for a nonspinning binary is nine. This Gaussian approximation is valid in the high SNR limit in which the square root of the diagonal entries of the corresponding covariance matrix correspond to lower bounds on the errors on each parameter. We do not discuss the details of this formalism, which is discussed in detail in \cite{CF94} in the context of GW data analysis. 

For all three LIGO detectors (L, H, I) we use the sensitivity curve labelled as 
{\it Zero Det, High P} and can be found in \cite{ligopsd}. For the KAGRA detector we use the curve labelled as 
{\it VRSE(B)} and can be found at the page \cite{kagrapsd} whereas the advanced Virgo noise 
can be found at the advanced Virgo project home page \cite{avirgo}.

\begin{figure}[ht]
\includegraphics[scale=0.5]{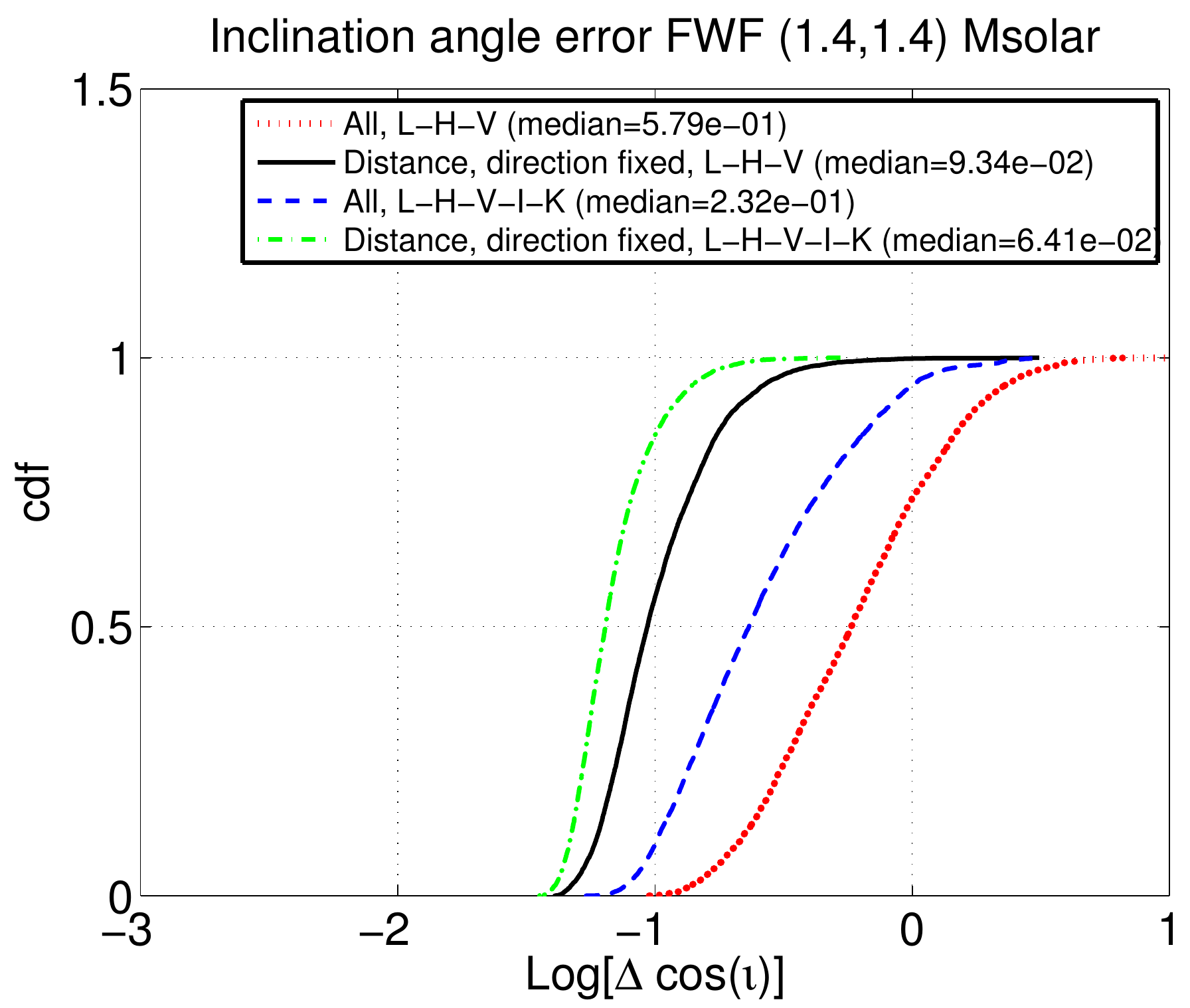}
\includegraphics[scale=0.5]{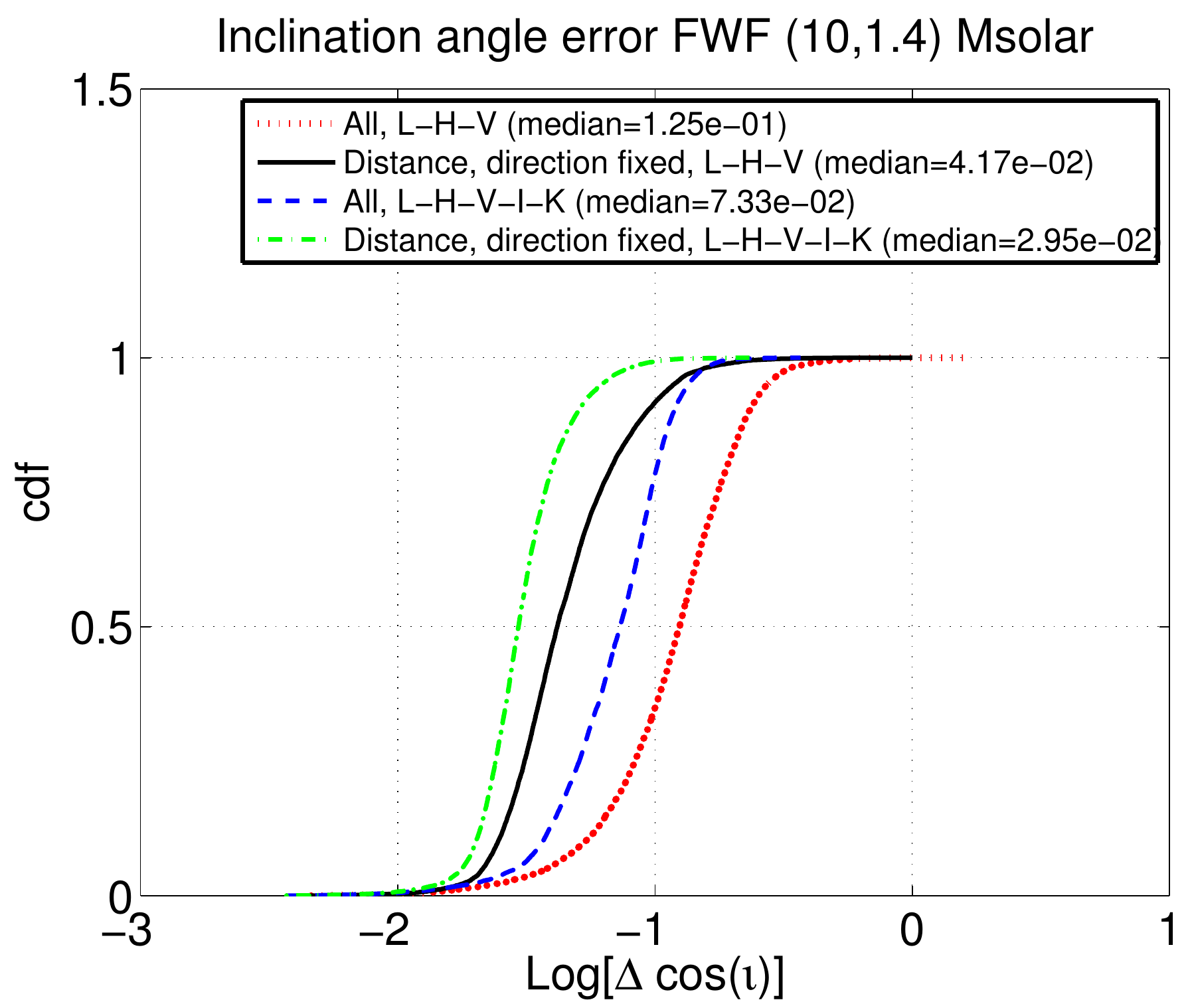}
\caption{Fisher matrix predictions for the accuracies of the estimation of the inclination angle of the binary (same as the viewing angle of a SGRB if it is associated with a binary merger) with a world-wide network of five GW detectors
for a double NS binary (top) of $1.4+1.4M_{\odot}$ and NS-BH binary of 
$1.4+10M_{\odot}$ (bottom). Distance to the binary is assumed to be 200 Mpc. Legends correspond to the cases where no EM information is available, and the SGRB is three dimensionally localized (distance and direction known) electromagnetically. The histogram is obtained using a population of binaries which are uniformly distributed and oriented in the sky.
}\label{fig:errors}
\end{figure}

Figure~\ref{fig:errors} displays the projected accuracies with which the errors on the cosine of the inclination angle may be estimated with three and five GW detector configurations by observing a DNS system of $2.8M_{\odot}$ and NS-BH system of $1.4+10M_{\odot}$, both of which are the  candidate progenitors of SGRB central engine.  The histograms correspond to a uniform source distribution over the sky surface, obtained by randomly choosing the source location, inclinations and polarization angles. However, all the sources have been kept at a fixed distance of 200 Mpc, which is approximately the average distance up to which DNS systems can be seen by GW detectors. For each realization we perform a Fisher matrix based error calculation where $\{{\cal M}, \eta, t_c, \phi_c, D_L, \iota,\theta,\phi,\psi \}$ are treated as independent parameters. ${\cal M}$ is a mass parameter, usually called {\it chirp mass}, that appears naturally in the gravitational waveform and is defined as ${\cal M}=M\eta^{3/5}$. $t_c$ and $\phi_c$ are kinematical parameters related to the time of arrival of the GW signal at the detector and the corresponding phase. We make a histogram of the distribution of the errors in  $\cos \iota$ for various astrophysical scenarios detailed below. (We quote here only the errors on the inclination angle. The errors associated with other parameters will be reported in detail elsewhere~\cite{TMPA2013}.)

Following Seto~\cite{Seto07}, we consider three astrophysical scenarios. (i) The GW signal is detected from a DNS or NS-BH binary, but no electromagnetic signals are seen. 
(ii) An EM observation is made in  with a GW signal where the EM observations provide the coordinates of the source direction and (iii) EM observations provide the direction to the source and, say, an optical counterpart gives the redshift estimation. In this case, using a model of cosmology, one can obtain the luminosity distance to the source from the observed redshift~\footnote{ The uncertainty in the cosmological model can introduce some errors in the redshift estimation which will have to be accounted.}. Hence, one can drop $\{\theta,\phi,D_L\}$ from the list of parameters to be estimated from GW observations. This is referred to as the 3D localized case in this paper.

We also compare the results using a GW detector network with (i) 3 advanced detectors consisting of 2 aLIGOs (Hanford and Livingston in the U.S.) and one advanced Virgo (near Pisa, Italy) and (ii) 5 detector configuration with 2 aLIGOs (L,H), advanced Virgo (V) and a Japanese detector KAGRA (K) along with a detector proposed to be installed in India, called LIGO-India (I).
 
Figure~\ref{fig:errors} shows the cumulative histogram of the errors in $\cos \iota$ for the two astrophysical scenarios (3D localized and no EM information) and for the population of sources discussed earlier.  Listed below are the important conclusions that can be drawn from the figure.
\begin{enumerate}
\item Typical $\Delta \cos\iota \sim 10^{-1}-10^{-2}$ which is about a few percent. For instance the inclination angle of a 3D localized source using a 5 detector configuration ( dash-dot/green curve) can be estimated with an error $\Delta\cos\iota \leq 10^{-1}$ for about 90\% of the injected DNS sources and $\Delta\cos\iota \leq  5\times 10^{-2}$ for 90\% of the injected NS-BH systems at 200 Mpc. These errors scale linearly with the luminosity distance. (See Fig.~\ref{fig:summary}.)
\item In the 3D localized case, when EM observations provide the information about the sky location and distance, there is a definite improvement in the estimation of the inclination angle for both the 3 detector and 5 detector cases, though it is less significant for a 5 detector network. This is not surprising since one will be fitting the data with 6 parameters as opposed to the original 9 parameters and this leads to improved estimation of the remaining parameters which includes the inclination angle. We have verified that the improvement in the inclination angle estimate is mostly influenced by the luminosity distance estimation from EM observations, as it  completely breaks the degeneracy between the inclination angle and the luminosity distance, leading to improved estimation of $\cos\iota$.
\item Adding more detectors, with distinct orientations, to the network indeed improves the estimation of 
the inclination angle.  This is mainly due to the increased sensitivity of a network (due to a larger coherent SNR) with larger number of detectors. However, partially, the improvement comes from the fact that the addition of more detectors resolves the 
degeneracy between various angular parameters, which, in turn, improves the inclination angle measurements.
\item The errors for the NS-BH system are smaller than the DNS system.  This is because, 
in the gravitational waveform, contributions from some of the harmonics (odd ones) vanish for 
symmetric systems (such as the DNS system) as these contributions are proportional to the 
asymmetry of the system characterized by the difference mass ratio parameter ($|m_1-m_2|/m$). 
Hence, for equal mass systems, there is less structure and information in the waveform as compared to an unequal mass system leading to worse estimation of the errors for equal mass systems.

\end{enumerate}
A quick summary of the median of the errors on $\iota$ for the 3 detector, 4 detector and 5 detector networks are  given in Table 1.  The errors are compared for the three cases: unlocalized, 2D localized, and 3D localized bursts. 

 The best constraints are for the 3D localized case 
and the worst case, naturally, is when no prior information is available from EM observations. The GW measurements would give tighter constraints on the inclination angle if the progenitor is a NS-BH system as opposed to a DNS system. As expected, an increase in the number of GW detectors helps tremendously in reducing the errors
and with the 5 detector network, even for the DNS progenitor case the errors become interesting.
\begin{table}
\begin{tabular}{|c|c|c|c|}
\hline
Network & No EM information& Direction known & 3D localized\\
\hline
LHV & 9.3 (41.5)& 8.3 (34.4) & 3.3 (8.6)\\
LHVK & 7.1 (24) & 6.5 (21.0) & 2.7 (6.4)\\
LHVKI & 5.8 (15.5) & 5.5 (14.3) & 2.2 (5.1)\\
\hline
\end{tabular}
\caption{Median $\Delta \iota$ in degrees for different detector network configurations for the scenarios where one does not make use of any EM information and the case where one assumes the location as well as the distance to the GRB is known electromagnetically (3D localized). The quantities quoted inside the brackets are for BNS progenitors of $2\times1.4M_{\odot}$ whereas the ones outside the bracket correspond to NS-BH progenitors $10+1.4M_{\odot}$.}
\end{table}

We would like to remind the readers again that the errors presented here are for a 
luminosity distance of 200 Mpc. The errors would scale linearly with an increase in the
luminosity distance (see Fig.~\ref{fig:summary}). For DNS systems, if the binary merger is farther away, 
the SNR may be low and the GW signal may go undetected. On the other hand observation of a NS-BH system, which can be seen farther away, might measure inclination angle with a reasonable accuracy though the errors will increase with distance. If $\Delta \iota_{200}$ is the error in measuring the inclination angle for a distance of 200 Mpc, the error for an arbitrary distance would scale as
\begin{equation}
\Delta \iota = \Delta \iota_{200}\,\left(\frac{D_L}{200 \;{\rm Mpc}}\right).
\end{equation}

In the next section, we discuss the implications of precisely estimating the inclination angle using joint GW-EM observations.
\section{Possible implications for SGRBs}\label{sec:GRB}
We now study the possible implications of inclination angle measurements 
from GW observations for the afterglow modelling of SGRBs.
We start with a quick summary of the important features of SGRB jets and light curves.
\subsection{Light curves of SGRB afterglows}
As we mentioned earlier, mergers of DNS or NS-BH binaries are strong candidates for progenitors of SGRBs (see \cite{ShGRBrevLee07} for a review on various progenitor models for SGRBs). In the simplest picture, the central BH (that is formed by the merger) accretes from the surrounding massive disk and powers the SGRB jet along the angular momentum axis of the BH. The SGRB jets are narrow with a typical semi-opening angle of the jet ($\theta_j$) to be of the order of $3-8$ degrees~\cite{SoderbergEtal2006,ShGRBJetBreak2011,FongEtal2013}. The associated isotropic luminosities of these bursts are $\simeq 10^{49}-10^{51}\;{\rm erg/s}$. While internal shocks are believed to be responsible for the prompt $\gamma$ ray emission, the afterglows at lower frequencies (from X-ray to radio) are produced by external shocks (due to the interaction of the jet with the circumburst medium). In many cases, the afterglow light curves (flux as a function of time since the burst trigger) show what is known as the {\it jet-break}~\cite{Rhoads99}. Jet break refers to the abrupt change in the power law index of the light curve and happens as the jet slows down with time and undergoes a lateral expansion.  The jet opening angle is related to the Lorentz factor of the jet by $\Gamma\simeq \theta_j^{-1}$. Since the SGRB emission is highly beamed, if we detect a SGRB, it means we are very close to the axis of the jet. Hence, usually GRB afterglow modelling is done assuming the GRB is {\it on-axis}. 
\subsection{Off-axis jets}
An on-axis GRB may be distinguished from an {\it off-axis} GRB, where the viewing angle, henceforth denoted as  $\theta_{v}$ (which is the angle between the line of sight and axis of the jet), is not equal to zero. Within the top hat or uniform jet model, very often used for interpreting GRB afterglows, the jet is assumed to be uniform  within a cone with sharp edges and has an opening angle $\theta_j$. Hence, if the GRB jet is slightly away from the jet axis but within the jet cone ($\theta_{v}\leq \theta_j$), the light curve would be identical to an on-axis GRB~\cite{GranotEtal2001,GranotEtalOffAxis2002} (see Fig.~1 of Ref.~\cite{GranotEtal2005} for the jet geometry).

But there may be cases where the observer's line of sight lies outside the cone of the jet ($\theta_{v} > \theta_j$). Such a jet is referred to as an off-axis jet whose light curves depend on the viewing angle~\cite{WLoffaxis99}. For such off-axis jets, the observed light curve will be faint initially as the beamed emission is not directed towards the observer. As the jet slows down and becomes wider, at some stage the observer's line of sight enters into the cone of the jet ($\Gamma^{-1}\sim \theta_{v}$), the light curve  peaks after which it decays like an on-axis ($\theta_{v}\simeq 0$) jet. Though there have been a few long GRBs which required one to invoke off-axis geometry to explain the afterglow spectrum (see, for instance, Refs.~\cite{GranotEtalOffAxis2002,Offaxis2005}), no such light curves have been reported so far for SGRBs yet. But there is no reason why SGRB jets are all on-axis. Hence, off-axis SGRB events are natural to occur.

If there is a GW event associated with a SGRB, then the $\theta_{v}$ of the GRB is nothing but the inclination angle of the binary which we discussed in detail in the earlier sections. Since GW observations would estimate the inclination angle very accurately, the afterglow modelling benefits from this information while interpreting the jet. 
\subsection{Implications of $\theta_{v}$ measurement for understanding off-axis jets}
From the discussion above, it is clear that the most important area where GW measurement of the viewing angle would make the maximum impact is in understanding the
geometry and energetics of off-axis jets in the case of SGRBs. We discuss below four specific scenarios where GW measurements could make an impact. Given the various uncertainties involved in the modelling, we will be qualitative in our discussions.

\subsubsection{Model independent constraints on the parameters \\of the off-axis jet model:}
One of the striking features of an off-axis GRB is the initial rise in the light curve which peaks at $t_{\rm p}$. This is related to the jet break time $t_j$ by~\cite{NakarOrphan07}
\begin{equation}
t_{\rm p} = A \left(\frac{\theta_{v}}{\theta_j}\right)^2\;t_j,
\end{equation}
where $A$ is an arbitrary constant roughly of the order of unity. Different models
predict values of $A$ close to unity on either side, depending on whether the peak flux is seen when the jet opening angle is before $\theta_{v}$, at that value or  after it. 

The ideal scenario would be the one where $t_j>t_p$, in which case EM observations are expected to measure both $t_j$ and $t_p$. When complemented with a GW measurement of the viewing angle, $\{t_j,t_p,\theta_v\}$ are quantities that can be directly inferred from EM/GW observations in a model independent way. Both $A$ and $\theta_j$ are usually inferred based on a model of the dynamics of the jet and a model of the circumburst medium, respectively. The observed $\{t_j,t_p,\theta_v\}$ can constrain the SGRB in the $A-\theta_j$ space independently of any model. Such a constraint can be very useful to understand the dynamics of the jet as well as the quantities related to the circumburst medium.

The other scenario is when $t_p>t_j$ when jet break is not observed electromagnetically. But knowledge of viewing angle from GW observations would help
the fitting of the afterglow light curve within any model, as it reduces the number parameters in the fit.

In short, an independent estimation of $\theta_{v}$ from GW observations  may be of great use in the afterglow modelling in the SGRB-GW era.
\subsubsection{Searches for Orphan afterglows of SGRBs}
The viewing angle information would also prove to be useful to understand the `orphan afterglows', where one does not observe the prompt gamma ray emission but discovers the afterglow serendipitously. The most popular model for such orphan afterglows is the scenario where the prompt emission is beamed so much away from the observer that the observer does not see it. But as the jet slows down and laterally expands, afterglow emission in lower frequency bands 
may be observed depending on the viewing angle of the jet. Though there are not any orphan afterglows detected yet even in the case of long GRBs, this is an interesting prospect especially if a GW signal is detected from the binary merger.

The GW measurement of the viewing angle may explain the non-observation of prompt emission from the SGRB. 
The GW observations may localize the compact binary within a  few square degrees for NS-BH systems and a few tens of square degrees for  DNS systems~\cite{Fairhurst2010,TMPA2013}.  More precisely, a 5-detector network of GW detectors (LHVKI) could be able to localize a DNS(NS-BH) systems within a 95\% confidence region of about 10 (5) square degrees~\cite{TMPA2013} at a distance of 200 Mpc. The corresponding numbers for the DNS and NS-BH systems with the LHV network are about 40 and 20 square degrees, respectively.
It may be possible to survey the corresponding patch of the sky after a time which can be roughly estimated from the viewing angle measurement and considering other
parameters to be that of typical SGRBs. Such a GW-triggered observation of a SGRB afterglow is among the brightest prospects of GW-SGRB synergy.  The data analysis challenge associated with the follow up of an orphan afterglow is discussed in detail in Ref.~\cite{GhoshBose2013}. A comprehensive study of the detectability of orphan radio afterglows from SGRBs was recently carried out in Ref.~\cite{Feng2014}.
\subsubsection{Off-axis geometry and $E_{\rm iso}-E_{\rm peak}$ relation for SGRBs}
 The tight correlation between the peak energy  of the GRB ($E_p$) in the cosmological rest frame and the isotropic total  radiated energy $E_{\rm iso}$ (also referred as $E_{rad}$ in the literature) of the burst (known as the {\it Amati relations}) has been an interesting feature of the GRBs ever since it was first reported~\cite{AmatiEtal2002,Eiso-Epeak2002b} in the case of long GRBs. In addition to shedding light on the nature of these bursts, such an empirical relation is helpful in deriving the (pseudo) redshifts of the GRBs whose redshifts are not directly known.
There have been investigations about a similar empirical relation for SGRBs.
Zhang et al.~\cite{ZhangEtal2012} concluded, using the available sample of SGRBs with known redshifts, that SGRBs also obey a similar relation though different from the one obeyed by long GRBs (see, also, Ref.~\cite{Eiso-EpeakShGRB2012}). They argued that for the long GRBs the relation is $E_p=100\times\left(E_{\rm iso}/10^{52}\right)^{0.51}$ while for SGRBs it is $E_p=2455\times\left(E_{\rm iso}/10^{52}\right)^{0.59}$.  This indicates that though the progenitors of long and short GRBs are different, the energy dissipation mechanism is similar.

 Since most of the observed SGRBs are likely to be on-axis (due to the selection bias), the above empirical relation is constructed assuming  on-axis jets.
If the SGRBs are off-axis and $\theta_{v} > \theta_j$ then the gamma ray fluxes may be strongly suppressed, as we saw earlier. The isotropic and peak photon energies will now depend on the viewing angle and  will fall of much more rapidly  as $[\Gamma\left(\theta_{v}-\theta_j\right)]^{-6}$ and $[\Gamma\left(\theta_{v}-\theta_j\right)]^{-2}$, respectively~\cite{GranotEtalOffAxis2002}. As a result, off-axis SGRBs may show deviations from the $E_{\rm iso}-E_{\rm p}$ curve~\cite{Offaxis2005}. Thus an independent GW measurement of the viewing angle would prove to be useful to correct the energies to account for the off-axis geometry.
 
 GW observations could play a role in distinguishing the outliers which are due to the off-axis geometry from those which are genuine. A sample of SGRBs with GW associations may be ideal to put {\it Amati relation} to test.
Further, even in the absence of a direct redshift measurement of the GRB by electromagnetic means, an
associated GW observation would allow direct measurement of the luminosity distance, which can be converted to the redshift using a cosmological model. 
\subsubsection{Implications for Structured Jet models}
 An alternative to the top hat jet model is the {\it universal structured jet} model~\cite{MeszarosEtal98,KumarGranotUSJ2003} (also known as the {\it inhomogeneous jet} model). In this model, the jet outflow is  wide, and the jet shows an angular structure due to which the kinetic energy of electrons and the Lorentz factor depend on the viewing angle (as power laws). Hence, the light curve would look different depending on the viewing angle of the jet, though the structure of the jet is the same for all bursts. Evidently, the viewing angle plays a very important role in this case, and an estimation of this from GW observations would significantly assist the interpretations within this model. Though this model is not yet used in the case of SGRBs, the SGRB-GW synergy may make it possible in the future to interpret the SGRB afterglows in the framework of structured jet models.
\section{Caveats and future directions}\label{sec:caveats}
We list below some of the caveats of our work and possible future projects.
\begin{enumerate}
\item As mentioned earlier, we have used an amplitude-corrected {\it full} PN waveform for our calculations. We have modelled the binary components to be point particles and ignored the tidal contributions, due to the finite size of the NS~\cite{1PNTidal2011}. These will be more important towards the late stages of the inspiral. A more realistic way would be to use the results from the numerical relativity simulations of DNS systems~\cite{ShibataEtal05} and NS-BH systems~\cite{KiuchiNSBHSim09}, which would be more complete. There are various semi-analytical and numerical models of the merger based on the effective one body approach and numerical relativity simulations (see, for instance, Ref.~\cite{BiniEOB2012}), which may also be used in a future work to revise our estimates.
\item We have completely ignored the spin effects in our parameter estimation.
Compact binaries containing at least one BH may have significant spins, in which case a spin parameter should be added to the 9 dimensional parameter space we considered. The effect of spins in the our estimates will have to be looked into in a future project.  Further, our assumption that  the inclination angle of the binary is same as the viewing angle of the jet will hold only if the binaries have negligible spins, or spins which are aligned or anti-aligned with respect to the angular momentum axis of the binary.
\item Recently, using the formalism of Ref.~\cite{CutlerVallisneri07}, Favata~\cite{Favata2013} studied  the systematic effects in parameter estimation due to the neglect of higher order PN terms and orbital eccentricity, in addition to the tidal effects and spins discussed above. Only a detailed investigation can quantify such systematic effects in our waveform model.
\item Another caveat of this analysis is the use of the Fisher information matrix to estimate the errors associated with the inclination angle measurement, which is the most important result. Though the Fisher information matrix is an elegant semi-analytical method to estimate the errors in a parameter estimation problem, the method works well in the high SNR limit and has its own limitations in the low SNR regime ~\cite{CF94,Vallisneri07}.
 It will be interesting to revisit the problem using techniques such as Markov-Chain Monte Carlo to sample the likelihood surface and estimate the parameters which we plan to address in future work.
\item Lastly, the realistic chance of such a joint observation is uncertain and will require very good coordination of the GW interferometers and EM telescopes. However, very few such joint observations might themselves bring tremendous breakthroughs in the field.

\end{enumerate}
Note added: After this paper was submitted, two papers~\cite{Ryan2014,Zhang2014} appeared arguing, by analysing a large sample of {\it Swift} GRB light curves, that GRBs are very likely to be off-axis and emphasized the importance of including off-axis effects in modelling the light curves of both long and short GRBs.
\acknowledgments
 K.G.A. is extremely thankful to L. Resmi for  discussions on various aspects related to the GRB afterglow modelling and very useful comments on the manuscript. The authors thank D. Bhattacharya, A. Kamble, I. Mandel, K. Misra and B. S. Sathyaprakash,  for useful discussions. We would like to thank T. Nakamura for useful comments on an earlier version of the manuscript.  We are grateful to Ray Frey for detailed comments on the manuscript. 
The project used the octave based codes developed by Roby Chacko, a project assistant under the AP's SERC Fast-Track Scheme. 
K.G.A. and A.P. acknowledge the DST and JSPS Indo-Japan International Cooperative Programme for scientists and engineers (DST/INT/JSPS/P-127/11) for supporting visits to Osaka University, Japan. H.T. thanks JSPS and DST under the same Indo-Japan programme for their visits to IISER Thiruvananthapuram, India. K.G.A. thanks IISER Thiruvananthapuram for hospitality during different stages of this work.
C.K.M.'s work was supported by MPG-DST Max Planck Partner Group on Gravitational Waves.
H.T.'s work was also supported in part by Grant-in-Aid for Scientific Research (C) No. 23540309, 
Grant-in-Aid for Scientific Research (A) Grant No. 24244028, and 
Grant-in-Aid for Scientific Research on Innovative Areas Grant No. 24103005.
This work was also supported by JSPS Core-to-Core Program, A. Advanced
Research Networks.
 This is LIGO document P1400031.
\bibliographystyle{apsrev}
\bibliography{ref-list}
\end{document}